**Thermal light emission from monolayer MoS$_2$**

*Lukas Dobusch\*, Simone Schuler, Vasili Perebeinos, and Thomas Mueller\**

L. Dobusch, S. Schuler, Prof. T. Mueller
Vienna University of Technology, Institute of Photonics, Gußhausstraße 27-29, 1040 Vienna, Austria
E-mail: lukas.dobusch@tuwien.ac.at, thomas.mueller@tuwien.ac.at
Prof. V. Perebeinos
Skolkovo Institute of Science and Technology, 3 Nobel Street, Skolkovo, Moscow Region 143026, Russia

Because of their strong excitonic photoluminescence (PL)[1–4] and electroluminescence (EL)[5–9], together with an excellent electronic tunability[10], transition metal dichalcogenide (TMD) semiconductors are promising candidates for novel optoelectronic devices. In recent years, several concepts for light emission from two-dimensional (2D) materials have been demonstrated. Most of these concepts are based on the recombination of electrons and holes in a pn-junction, either along the lateral direction using split-gate geometries in combination with monolayer TMDs[5–7], or by precisely stacking different 2D semiconductors on top of each other, in order to fabricate vertical van der Waals heterostructures, working as light-emitting diodes (LEDs)[12,14]. Further, EL was also observed along the channel of ionic liquid gated field-effect transistors (FETs) under ambipolar carrier injection[15,16]. Another mechanism, which has been studied extensively in carbon nanotubes (CNTs)[17–20] and more recently also in graphene[21,22], is thermal light emission as a result of Joule heating. Although the resulting efficiencies are smaller than that of LEDs based on ambipolar electron-hole injection, these experiments provide valuable insights into microscopic processes, such as electron-phonon and phonon-phonon interactions, and the behavior of low-dimensional materials under strong bias in general.

Thermal light emission from a nanomaterial occurs at elevated electron temperatures, which lead to thermal population of exciton states and subsequent radiative recombination. High temperatures can be reached by minimizing heat dissipation in the vertical direction through the underlying substrate and along the lateral direction towards the contact electrodes, acting as heat sinks. Vertical heat dissipation can be greatly suppressed by using suspended materials in vacuum[17–19,21], while the heat flow in the lateral direction is determined by the material's thermal conductivity. Several theoretical[23–25] and experimental[26,27] studies have revealed low values for the lattice thermal conductivity of TMDs, ranging from 34 Wm$^{-1}$K$^{-1}$ to 131 Wm$^{-1}$K$^{-1}$ at 300 K in the case of monolayer $MoS_2$. The thermal conductivity of $MoS_2$ is thus about two orders of magnitude smaller than that of graphene[28] (~5000 Wm$^{-1}$K$^{-1}$) or CNTs[29] (~6600 Wm$^{-1}$K$^{-1}$). $WSe_2$, another 2D semiconductor, exhibits an even lower thermal conductivity of ~4 Wm$^{-1}$K$^{-1}$ in a monolayer crystal and even less in disordered films[30]. The low lateral heat dissipation makes $MoS_2$ and other TMDs interesting materials for thermoelectric applications[31–34] at the nanoscale. On the other hand, it may lead to strongly localized hot spots that limit the performance and reliability of electronic devices.

In this communication, we show that a $MoS_2$ monolayer sheet, freely suspended in vacuum over a distance of 150 nm, emits visible light as a result of Joule heating. Due to the poor transfer of heat to the contact electrodes, as well as the suppressed heat dissipation through the underlying substrate, the electron temperature can reach ~1500–1600 K. The resulting narrow-band light emission from thermally populated exciton states is spatially located to an only ~50 nm wide region in the center of the device. We employ a backgated n-channel field-effect transistor (nFET), operated in the resistive regime, i.e. with negative drain-source bias ($V_{DS} < 0$), in order to avoid channel pinch-off at the drain and thus achieve high electric fields over the trench (see Supporting Information). The onset of light emission is accompanied by a negative differential conductance (NDC) regime, indicating severe self-heating and scattering of electrons by optical (OP) phonons[19,21].

A schematic sketch of our device is presented in Figure 1a. Figure 1b shows the corresponding optical micrograph. The MoS$_2$ monolayer flake is freely suspended over a predefined trench with a width of $d = 150$ nm (see Experimental section for fabrication details). MoS$_2$ was chosen because it allows to drive sufficiently large currents for thermal emission; we note that the device concept can be applied to any other 2D semiconductor. The distance between the contact electrodes ($L_{Channel} = 4$ µm) is chosen sufficiently large to clearly determine the spatial location of the resulting light emission and distinguish it from emission from the contacts, which was previously reported by Sundaram *et al.*[35]. As shown in the inset in Figure 1a, our calculation suggests that (under strong bias) a large fraction of the applied drain-source voltage $V_{DS}$ drops at the trench, i.e. $V_{trench}/V_{DS} \approx 0.5$, even though $d/L_{Channel} \approx 0.04$, allowing effective heating in the suspended MoS$_2$. The scanning electron micrograph in Figure 1c shows the suspended monolayer MoS$_2$, where neither bendings or strain, nor cracks could be observed. A MoS$_2$ thickness of one layer was confirmed by Raman measurements[36], performed before the flake transfer (Figure 1f). The Si substrate is used as a backgate with a $t_{SiN} = 100$ nm thick high-k gate dielectric (Si$_3$N$_4$). Transfer- and output-characteristics were recorded before optical measurements under light emitting operation. From the transfer-characteristic, shown in Figure 1d, a field-effect mobility of $\mu \approx 17.5$ cm$^2$V$^{-1}$s$^{-1}$ at a substrate temperature of $T_0 = 80$ K is extracted using $\mu = (L/W) \times (1/c_{SiN}) \times (dG/dV_{BG})$. Here, $G = I_{DS}/V_{DS}$ is the electrical conductance, $L$ and $W$ are the MoS$_2$ channel length and width, respectively, and $c_{SiN} = \varepsilon_{SiN}\varepsilon_0/t_{SiN}$ is the gate oxide capacitance per unit area. We use an effective channel length in our calculations, given by $L = L_{Channel} + d(\varepsilon_{SiN} - 1)$, including carrier density modulation described by the parallel capacitor model, and $\varepsilon_{SiN} = 7$ ($L_{Channel} = 4$ µm and $W = 4.5$ µm). Short-channel effects are neglected, because of the short screening length[37,38] $\lambda = \sqrt{t_{SiN}\, t_{MoS2}\varepsilon_{MoS2}/\varepsilon_{SiN}} \approx 6$ nm ($t_{MoS2} \approx 6.5$ Å and $\varepsilon_{MoS2} \approx 4$ denote the thickness and dielectric constant[39] of monolayer MoS$_2$, respectively), and thus $\lambda \ll d$. The extracted mobility is in agreement with previous studies on similar devices. At low $V_{DS}$, self-heating effects can be neglected, hence $T_{el} = T_0$ with $T_{el}$ being the electron temperature.

If the nFET is operated in its on-state ($V_{BG} > V_T$, with $V_T$ denoting the transistor threshold voltage), light emission sets in above $V_{DS} \lesssim -15$ V and rapidly increases with more negative $V_{DS}$. Optical micrographs reveal that the origin of light emission is located at the suspended region of the flake (Figure 2a), where heat dissipation through the underlying substrate is eliminated. Further evidence is given by considering the temperature profiles shown in Figure 4a, which we obtained by device simulations. Figure 2b shows the emitted light intensity (upper plot) and corresponding current $I_{DS}$ (lower plot) versus applied bias voltage $V_{DS}$. At large $V_{DS}$, the device enters the NDC regime, which is a clear indication of self-heating and electron scattering by OP phonons or charge transport through more than one conduction band valley due to band effects under high biasing. A strong temperature dependence of the mobility in monolayer MoS$_2$ due to OP scattering was theoretically predicted[40] and experimentally[41] observed to occur at temperatures above ~200 K. The influence of self-heating upon device operation was also investigated previously by high-field transport measurements on MoS$_2$ transistors[42].

Such heating is caused by Joule power dissipation, $P \approx V_{trench} \times I_{DS}$, where hot electrons emit OP phonons, which then decay through anharmonic coupling into acoustic (AC) modes that carry the heat away from the trench[19–21]. The OP phonons give rise to strong scattering of electrons[40]. We point out that, whereas typically $T_{el} \approx T_{OP}$, AC and OP phonon modes do not necessarily have to be in equilibrium. In graphene and CNTs, for example, $T_{OP} > T_{AC}$, caused by a bottleneck in the decay of OP phonons into lower energy AC phonons. This imbalance can be described by the non-equilibrium phonon coefficient[19–21] $\alpha > 0$ (graphene: ~0.4; CNTs: ~2.4) with $T_{OP} = T_{AC} + \alpha (T_{AC} - T_0)$, where $T_{OP}$ and $T_{AC}$ are the OP and AC phonon temperatures, respectively. Unlike in graphene[43] and CNTs, phonon anharmonicity is much stronger[44] in MoS$_2$, which results in a smaller thermal conductivity[45] and more efficient energy exchange between the phonon branches. Moreover, electron-phonon coupling to AC and OP phonons does not vary by two orders of magnitude[40] in MoS$_2$. Therefore, a similar population of both AC and OP phonon branches is expected under current flow conditions, i.e. $T_{OP} \approx T_{AC}$.

Under light emitting operation, the high electron temperature $T_{el}$ leads to an appreciable thermal population of exciton states and visible light emission due to radiative electron-hole pair recombination. We note that, in principle, thermal light emission can also occur in bulk or few-layer TMDs with an indirect bandgap[46], provided that the direct gap is thermally populated. However, due to lower power density in the channel, higher currents are then required to achieve same $T_{el}$ as in monolayers, rendering such devices less efficient. Figure 2c presents thermal emission spectra for $V_{DS}$ ranging from -17 V to -20 V and $V_{BG}$ = 18 V, plotted on a logarithmic scale, where the well pronounced peak around ~ 1.7 eV corresponds to the A-exciton. The second (weaker) peak at higher energy, referred to as B-exciton, arises from the spin-orbit splitting of the valence band and corresponds to the transition from the lower spin level of the valence band. While the spin degeneracy results in a splitting of ~ 160 meV at low temperature[47], this value is assumed to increase with temperature since the redshift of the B-exciton with temperature is reported to be less pronounced than the A-exciton energy shift[48]. Compared to the PL spectra, shown in Figure 1e, the thermal emission spectra are considerably broader and shifted in energy.

In order to extract the bias-dependent electron temperatures $T_{el}$, we fit the spectra (dashed lines in Figure 2c) by Planck's law

$$I(\nu, T_{el}) = \frac{2h\nu^3}{c^2}\left(\exp\left(\frac{h\nu}{k_B T_{el}}\right) - 1\right)^{-1} \epsilon(\nu), \qquad (1)$$

modified by a frequency dependent emissivity $\epsilon(\nu)$. Here, $h$ is Planck's constant, $c$ the speed of light, and $\nu = E/h$ the frequency. The emissivity of a material equals its absorptivity, which we model by two Gaussians for the A and B-excitons in $MoS_2$, respectively, as described in the Supporting Information. At $V_{BG} = 18$ V we find a maximum electron temperature of $T_{el} \approx 1597$ K (Figure 2c). Spectra for other gate voltages can be found in the Supporting Information. By plotting the emission intensity versus the extracted $T_{el}$ (Figure 2d, symbols), we find good agreement with the theoretically expected behavior (dashed line) according to equation (1).

For the observed redshift, several mechanisms could, in principle, be responsible. The gate bias increases the doping level in the $MoS_2$ sheet, which might screen the Coulomb interaction and lead to both a renormalization of the bandgap and a modification of the exciton binding energy[49,50]. Gating might also strain the material over the trench, resulting in a bandgap change. However, as depicted in the inset in Figure 2d, we do not find any appreciable dependence on $V_{BG}$. Alternatively, the lateral electric field over the trench may induce a Stark-shift, but previous work suggests that the shift would be almost two orders of magnitude smaller than that observed in our experiment[51,52]. Instead, the shift can mostly be attributed to lattice heating. When plotting the A-exciton position versus $T_{el}$ (inset in Figure 2d), we find a linear relation with a slope of -0.20 meVK$^{-1}$. In order to compare this with the PL's spectral shift at elevated temperatures we use Varshni's relation[53] and extrapolate from PL measurements taken below room temperature (Figure 1e). Varshni's relation describes the reduction of the bandgap energy $E_g$ with increasing lattice temperature given by $E_g(T) = E_0 - aT^2/(T+b)$, with $a$ and $b$ being fitting parameters[54]. We find a redshift with -0.25 ± 0.03 meVK$^{-1}$ (see Supporting Information), similar to the value extracted from the thermal emission measurements.

In order to obtain further insight into the experimental data, we performed one-dimensional numerical device simulations along the transport direction. In doing so, we find the temperature distribution by solving the heat conduction equation

$$\kappa A \frac{\partial^2 T_{AC}}{\partial x^2} + p' - g(T_{AC} - T_0) = 0, \qquad (2)$$

where $\kappa$ is the thermal conductivity of MoS$_2$, $p' = I_{DS} \, dV/dx$ is the Joule heating per unit length, $A = W t_{MoS_2}$ is the MoS$_2$ cross-sectional area, and $g$ is the heat loss per unit length due to heat dissipation through the underlying substrate. $g$ is zero at the suspended region of the flake, while $g_{supported}$ considers the interfacial heat conductance of MoS$_2$ and the thermal resistance of the Si$_3$N$_4$ dielectric ($g_{supported,80K} \approx 9.7$ Wm$^{-1}$K$^{-1}$). A detailed description is given in the Supporting Information. The Si substrate is assumed to act as heat sink at temperature $T_0$. Radiation losses have a negligible effect on the temperature distribution and are thus not included in our simulation.

In a first step, we adapted our model (see Experimental section and Supporting Information) in order to reproduce the measured IVs under low biasing condition (i.e. without self-heating), using the characteristics obtained from electrical measurements. We model the carrier mobility as $\mu^{-1} = \mu_{imp}^{-1} + \mu_{OP}^{-1}$ including scattering by charged impurities and phonon scattering with $\mu_{OP} \sim T_{OP}^{-1.4}$, taken from the work by Radisavljevic et al.[41], with $\mu_{80K} = 17.5$ cm$^2$V$^{-1}$s$^{-1}$ at $T = 80$ K (see Figure 1d). We then proceed by solving self-consistently thermal and electrical transport, where we use $\kappa = \kappa_{300}(300 \text{ K}/T_{AC})^\beta$ with $\beta = 1$ for the temperature dependent thermal conductivity, consistent with Umklapp scattering[25]. $\kappa_{300} = 84$ Wm$^{-1}$K$^{-1}$ is taken from reference 27. Despite the large thermal gradient at the trench (up to ~10 K nm$^{-1}$), thermoelectric effects[31,32] do not substantially influence the electrical transport and are neglected (see Supporting Information). Our calculations allow an estimation of the maximum $T_{el} \approx 1600$ K at $V_{BG} = 18$ V, which is in agreement with the value extracted from the thermal emission spectra. Figure 3 shows experimental (a & b) and calculated (c & d) current- and intensity-maps, showing again good agreement. The NDC regime is well pronounced in the upper left corner at large $V_{DS}$ and $V_{BG}$ (Figure 3a & c), which is where light emission is observed (Figure 3b & d).

The calculated temperature distribution, shown in Figure 4a, indicates strong heat confinement over the trench, enabled by the low thermal conductivity of monolayer MoS$_2$ and, consequently, a short transfer length $L_T = \sqrt{\kappa A/g} < 100$ nm of heat to the substrate. This is in contrast to CNTs or graphene, where the emission occurs over a micrometer distance because of the much higher thermal conductivity $\kappa$. From the simulated temperature profile and using equation (1), we estimate that the thermal emission stems from an only $d_{rad} \approx 51$ nm (full-width-at-half-maximum) wide region in the middle of the trench, as illustrated in the inset in Figure 4a. An estimation of upper and lower bounds for $d_{rad}$ can be found in the Supporting Information.

Using a calibrated photodiode, we detect an emissive power of $P_{rad} \approx 1.3$ pW at $V_{BG} = 18$ V and $V_{DS} = -20$ V, or, when normalized to unit area and solid angle, $P'_{rad} = P_{rad}/(d_{rad}W\Omega_{coll}) \approx 11$ Wm$^{-2}$sr$^{-1}$, where $\Omega_{coll} \approx 0.52$ sr is the collection angle of our microscope objective (see Experimental section). By integrating equation (1) over $\nu$ and adjusting $T_{el}$ such that the experimental result for $P'_{rad}$ is reproduced, we obtain $T_{el} \approx 1520$ K. This value deviates by only ~5 % from the result of the spectral fitting procedure above and is less susceptible to the A/B-exciton ratio of the emissivity, because most of the emission stems from the A-exciton. We thus conclude that maximum $T_{el}$ values are in the range ~1500–1600 K.

Figure 4b shows possible reasons for light emission from our device, pointing out two additional mechanisms besides thermal emission. Under ambipolar carrier injection, electrons and holes are injected simultaneously at opposite ends of the channel, thus enabling radiative recombination. In n-type MoS$_2$ such EL can occur in the region close to the hole injecting contact[15]. We can therefore exclude ambipolar carrier injection as reason for light emission from our device by considering the optical micrographs shown in Figure 2a, which reveal that the origin of light emission is located at the region of the trench. Another EL mechanism is impact excitation due to coulombic electron-electron interaction[55]. This process requires large electric fields in order to accelerate carriers to energies above a threshold, enabling electronic excitation across the bandgap. Thereby, the threshold electric field is determined by the optical bandgap energy $E_g$ and the electron mean free path $\lambda_{OP}$ with respect to OP phonon scattering[55]. The threshold electric field can be estimated to be $\mathcal{E}_{T,ie} \approx 1.5\, E_g/(q\lambda_{OP})$, where the factor of 1.5 stems from momentum conservation restrictions[56] and $q$ is the elementary charge. In monolayer $MoS_2$, $\mathcal{E}_{T,ie}$ can be estimated to ~200 V/μm, when a mean free path[40] of 14 nm and a bandgap[3] of 1.85 eV are assumed. We note that this value constitutes a lower bound, as $\lambda_{OP}$ decreases further with higher temperature. On the other hand, our calculations suggest that light emission sets in at much lower fields in the range of ~40 V/μm. Without self-heating this value decreases to ~16 V/μm, which allows us to rule out impact excitation as the underlying mechanism for electrically driven light emission.

In conclusion, we have demonstrated thermal light emission from monolayer $MoS_2$ at the nanoscale. The freely suspended $MoS_2$ sheet can reach temperatures as high as ~1500–1600 K and support electrical power densities of up to ~0.25 MWcm$^{-2}$ before sudden device failure occurs. The low oxygen partial pressure in our vacuum chamber allows the $MoS_2$ to reach such high temperatures without oxidation. Reports on the melting point of monolayer $MoS_2$ are lacking, but melting of bulk material[57] was found to arise at temperatures >1923 K. Our measurements are well described by Planck's law, even though the small size of the emission spot ($d_{rad} < hc/E_0$) places it outside the thermodynamic limit regime[58]. From a practical point of view, our results are also important with regard to the miniaturization of TMD-based electronic devices. The low thermal conductivity in TMD semiconductors, together with high thermal power dissipation per unit area, gives rise to high electron and lattice temperatures, which could impact device performance and reliability. Hot carriers can be injected into forbidden regions of the device, such as the gate oxide, where they may get trapped or cause interface states to be generated. Thermal management may thus be necessary to improve the reliability of TMD-based electronics. This is especially relevant for devices on substrates with low thermal conductivity, such as plastics used in flexible electronics. Moreover, the device concept could be applied in the mid-infrared spectral regime by replacing $MoS_2$ with a narrow-gap 2D semiconductor, such as e.g. $ZrSe_2$. We estimate that, because of the smaller optical band gap, thermal population of exciton states will then be more than two orders of magnitude more efficient than in the visible, resulting in efficient mid-infrared narrowband emission.

**Experimental Section**

*Device fabrication:* Device fabrication began with uniform deposition of a 100 nm thick silicon nitride ($Si_3N_4$) layer onto a doped silicon (Si) wafer by plasma-enhanced chemical vapor deposition (PECVD). Using standard electron-beam lithography, 150 nm wide trenches were patterned and subsequently etched by reactive ion etching (RIE) using $CHF_3$. In a second lithography step, Ti/Au (10 nm/100 nm) bonding pads were defined. $MoS_2$ was mechanically exfoliated from bulk (SPI Supplies) onto a stack of poly(methyl methacrylate)/polyacrylic acid (PMMA/PAA) deposited on top of a sacrificial Si wafer. The layer thickness of the two polymers was chosen such that $MoS_2$ monolayers could be identified by optical microscopy. In order to confirm the $MoS_2$ layer thickness, we performed Raman spectroscopy prior to the transfer of the flake (532 nm, 150 µW). For the transfer, the PAA layer was dissolved in water, thus releasing the PMMA layer (not water-soluble) with the monolayer flake. The PMMA layer was then placed on a transparent polydimethylsiloxane (PDMS) stamp, turned upside down, and positioned with micrometer-precision on the pre-fabricated substrate such that the trench fully separates the flake into two similar sized regions. Contact electrodes were patterned directly onto the transferred PMMA by electron-beam lithography and Ti/Au (2 nm/60 nm) evaporation.

*Characterization:* For device characterization, the sample was placed inside a cryostat (Oxford Instruments) and cooled down to 80 K under high vacuum ($10^{-6}$ mbar). The emitted light was collected by a 20× microscope objective (numerical aperture $NA = 0.4$) and fed into a grating spectrometer (Horiba iHR320), equipped with a silicon photodetector array (Horiba Symphony II). Electrical characteristics were recorded simultaneously with the optical measurements using sourcemeters (Keithley 2612A). For measurements of the radiated power, a photodiode with femto-Watt sensitivity was placed directly behind the objective lens. The solid collection angle was calculated from the numerical aperture according to $\Omega_{coll} = 2\pi(1 - \sqrt{1 - NA^2}) \approx 0.52$ sr.

*Device model:* IV-curves were calculated by $I_{DS} = V_{DS}/R$, with $R = 2R_C + L/(\sigma W)$, where $R_C$ is the contact resistance, modeled according to our findings from 4-point measurements. $\sigma = qn\mu$ is the electrical conductivity, with carrier mobility $\mu$ obtained from electrical characterization ($\mu_{80K} = 17.5$ cm$^2$V$^{-1}$s$^{-1}$), and carrier density $n$, calculated by the parallel-plate capacitor model using $n \approx (V_{BG} - V_T - \xi V_{Channel})c_{SiN}/q$ above threshold, with channel-potential $V_{Channel}$ and a constant bulk effect parameter $\xi = 0.59$ at the supported region ($c_{SiN} = 620$ µFm$^{-2}$), taking drain induced doping effects into account[41]. Note that $c_{trench} = c_{SiN}/\varepsilon_{r,SiN}$ at the suspended flake ($\varepsilon_{r,SiN} = 7$). We include scattering from charged impurities and OP phonon scattering for the temperature dependency of $\mu$ and an Umklapp scattering limited $\kappa$ (see main text). We solve the heat conduction equation (2) numerically in an iterative approach until the stepwise change in $T_{max}$ is less than 0.2 K. See Supporting Information for further details.

**Supporting Information**

Supporting Information is available from the Wiley Online Library or from the authors.

**Acknowledgements**

We thank Markus Glaser and Marco Furchi for technical assistance and acknowledge financial support by the European Union (grant agreement No. 696656 Graphene Flagship) and the Austrian Science Fund FWF (START Y 539-N16).

Received: ((will be filled in by the editorial staff))
Revised: ((will be filled in by the editorial staff))
Published online: ((will be filled in by the editorial staff))

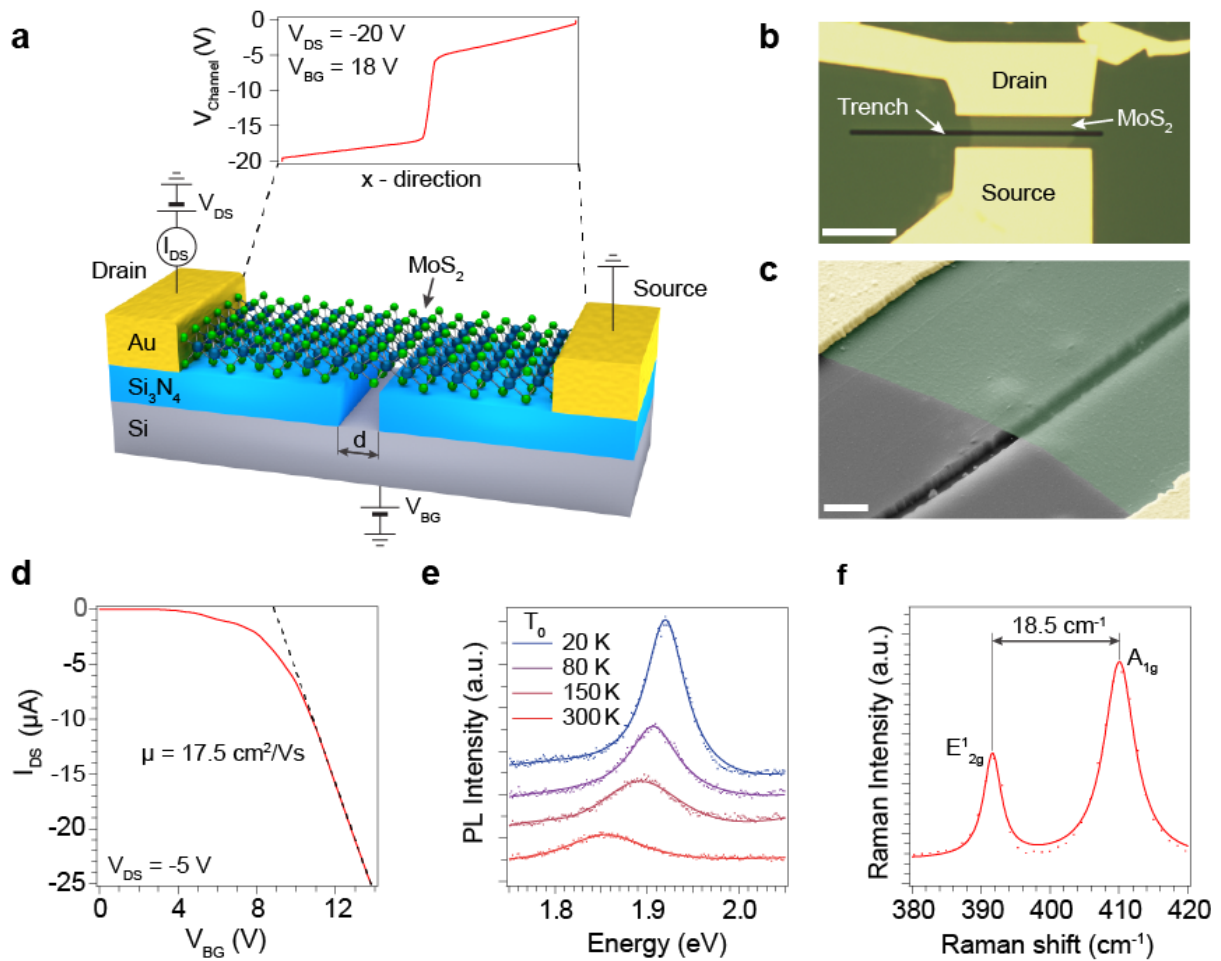

**Figure 1.** (a) Schematic sketch of the device structure with monolayer MoS$_2$ freely suspended over a trench (width $d$ = 150 nm). The MoS$_2$-flake is contacted by Ti/Au (2 nm/60 nm) electrodes, and separated from the Si substrate (used as backgate) by a 100 nm thick gate dielectric (Si$_3$N$_4$). The inset shows the calculated channel Potential $V_{Channel}$, when $V_{BG}$ = 18 V and $V_{DS}$ = -20 V. The small (~0.5 V) potential drops at $x = 0$ and $x = L$ are due to the contact resistance. (b) Optical and (c) scanning electron micrograph of a typical device. (scale bars: (a) 10 μm, (b) 400 nm) (d) Transfer characteristic taken at $V_{DS}$ = -5 V and $T_0$ = 80 K. ($W$ = 4.5 μm, $L$ = 4 μm) (e) Offset photoluminescence from monolayer MoS$_2$ for various temperatures, ranging from 20 K (blue) to 300 K (red). (f) Raman spectrum of monolayer MoS$_2$. The frequency difference of 18.5 cm$^{-1}$ between the E$^1_{2g}$ and A$_{1g}$ Raman modes indicates a thickness of one layer.

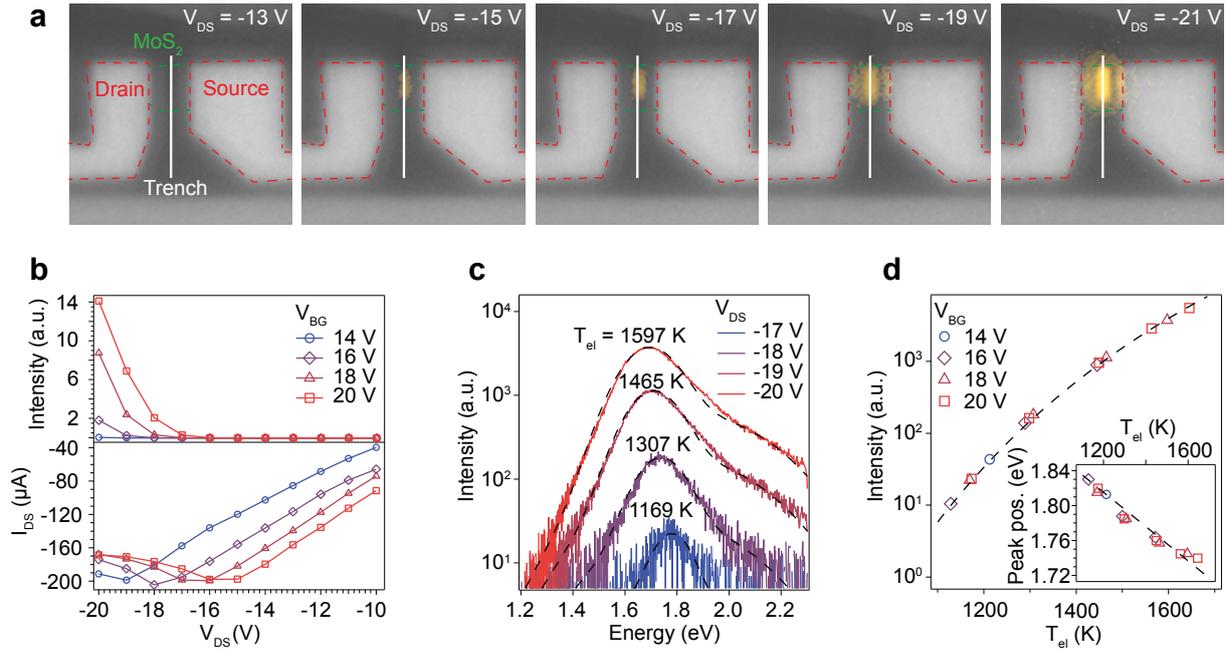

**Figure 2.** (a) Optical micrographs of the thermal light emitter for various biasing conditions, ranging from $V_{DS}$ = -13 V to -21 V (from left to right). Note that the origin of light emission is located at the suspended region of the MoS$_2$ flake. (b) Integrated thermal emission intensity (upper plot) and corresponding current $I_{DS}$ (lower plot) vs. bias voltage $V_{DS}$ for $V_{BG}$ ranging from 14 V to 20 V. The onset of light emission is in the negative differential conductance (NDC) regime under high biasing, indicating severe self-heating and scattering of electrons by hot OP phonons. (c) Measured thermal emission spectra at a logarithmic scale with $V_{BG}$ = 18 V and $V_{DS}$ ranging from -17 V to -20 V. The corresponding electron temperature $T_{el}$ can be extracted by using Planck's law, as indicated by the dashed lines. (d) Thermal light emission intensity plotted on a logarithmic scale as a function of $T_{el}$. The dashed line is a fit to the measurement data using Planck's law. The inset shows the temperature dependent shift of the A-exciton peak with $T_{el}$, where the dashed line is a linear fit.

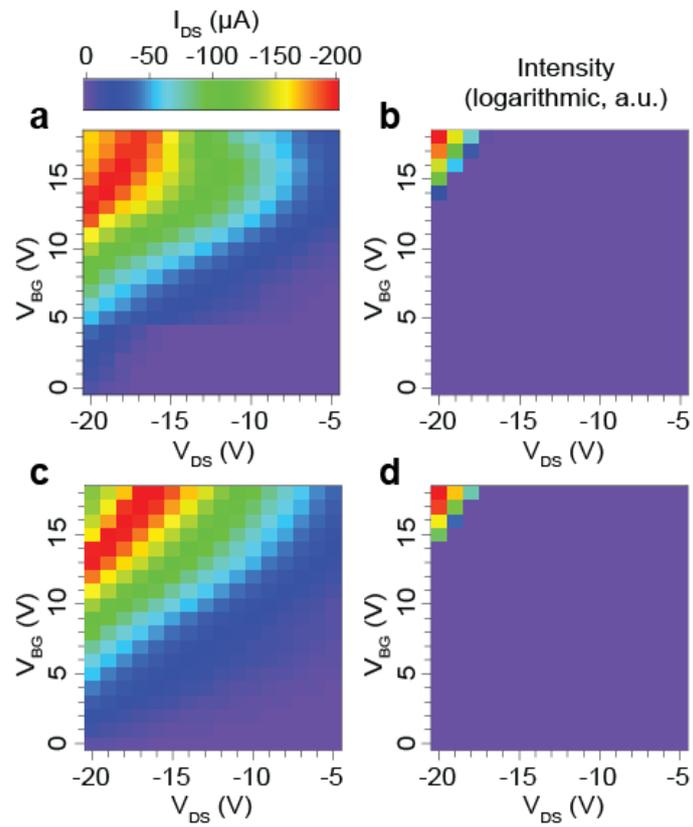

**Figure 3.** (a) Measured current-map and (b) corresponding emission intensity-map (logarithmic). Note that the onset of light emission is accompanied by a kink in the device's IV, indicating the NDC regime. (c) and (d) show the calculated equivalent.

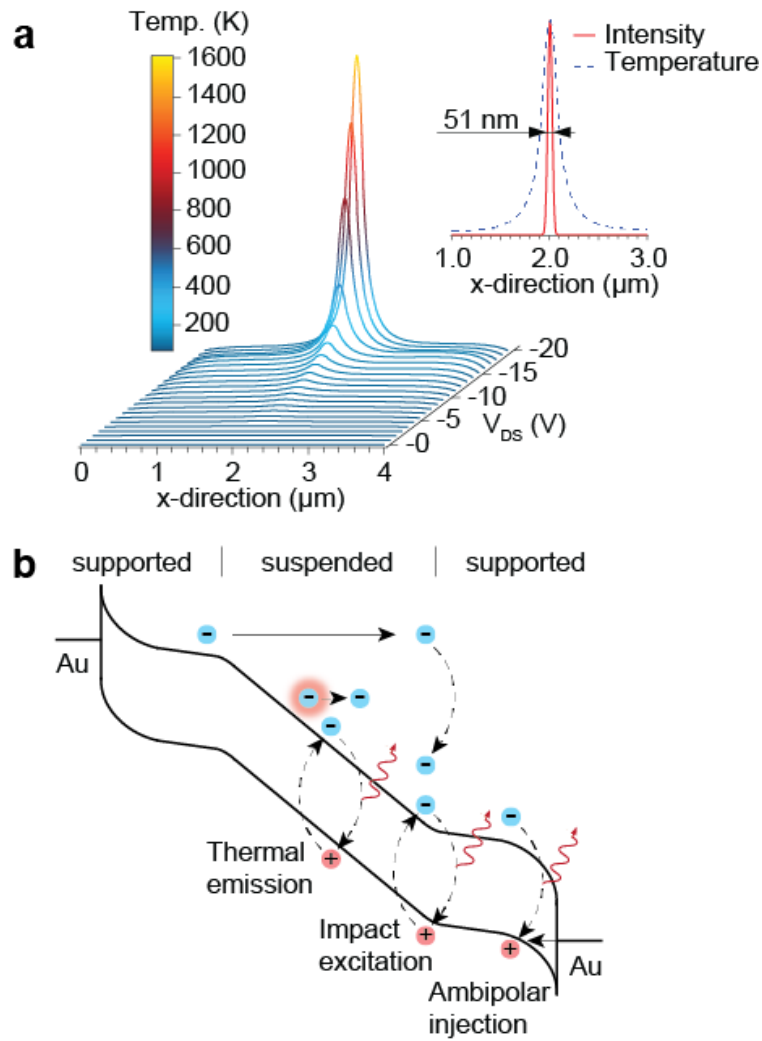

**Figure 4.** (a) Calculated distribution of the lattice temperature along the channel for $V_{DS}$ ranging from 0 V to -20 V and $V_{BG}$ = *18 V*. (b) Possible mechanisms for light emission. *Thermal light emission* occurs at the region of highest temperature (i.e. the center of the trench), due to thermal population of exciton states and subsequent radiative recombination. Large electric fields can accelerate carriers to energies sufficient to create excitons across the bandgap by electron-electron *impact excitation*. Under *ambipolar injection*, electrons and holes are simultaneously injected at opposite ends of the flake, which can then recombine radiatively.

# Supporting Information

**Thermal light emission from monolayer MoS$_2$**

*Lukas Dobusch\*, Simone Schuler, Vasili Perebeinos, and Thomas Mueller\**

**Contents:**

1.) Regimes of device operation

2.) Additional thermal emission spectra

3.) Varshni's relation

4.) Device modeling

5.) Influence of thermal gradient on potential

6.) Upper and lower bounds for $d_{rad}$

**1.) Regimes of device operation**

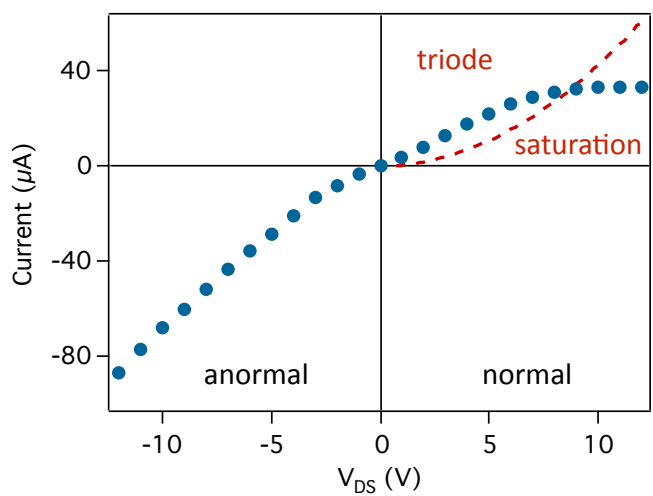

**Figure S1.** IV-characteristic, showing different regimes of transistor operation (without self-heating). For thermal light emission, the device is operated in the anormal (resistive) regime.

## 2.) Additional thermal emission spectra

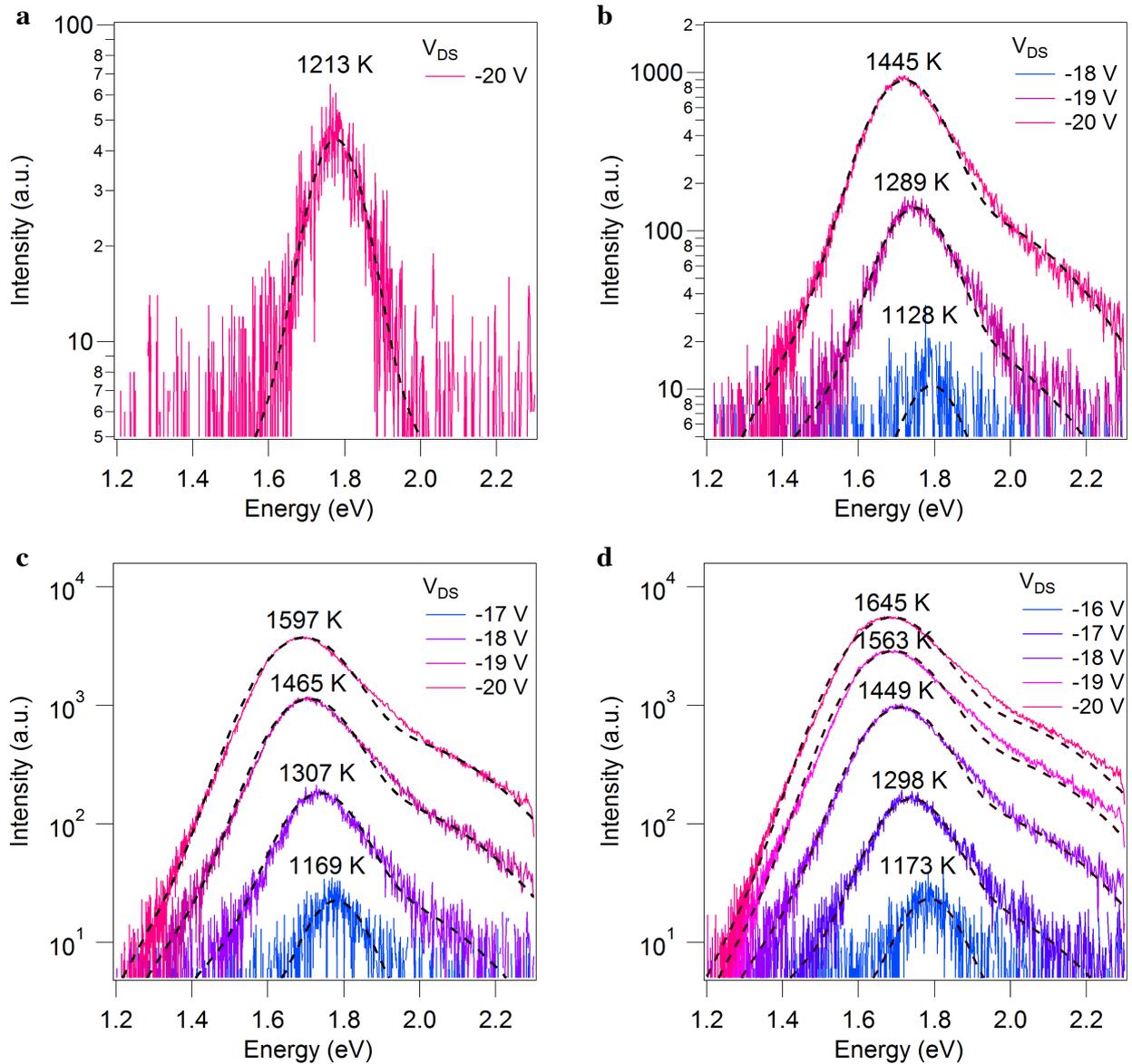

**Figure S2.** Measured thermal emission spectra for $V_{BG}$ = 14 V (a), 16 V (b), 18 V (c) and 20 V (d). The dashed lines indicate the fit that was used for extracting the corresponding electron temperatures, as discussed in the main article. The obtained electron temperatures are shown in the plots.

Figure S2 shows additional thermal emission spectra for the device presented in the main article for different biasing conditions. The corresponding IV-curves are presented in Figure 2b in the main text. In order to extract the corresponding electron temperatures, emission spectra were fitted after equation (1) from the main text. Here we model the emissivity $\epsilon$, which equals the materials absorptivity, by two Gaussians with same amplitude for the A- and B-excitons, and use a peak value of $\epsilon_{max} = 0.08$ [1]. In the fitting procedure we vary $T_{el}$, which directly influences the emission intensity, the peak position (inset in Figure 2d in the main text) and spectral width of the A-exciton. The used emissivity spectrum for $T = 1600$ K is shown in Figure S3.

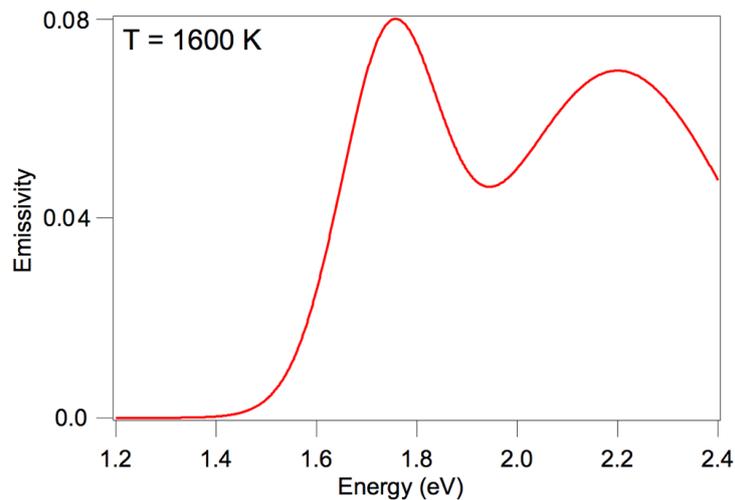

**Figure S3.** Emissivity model used for the fitting procedure after equation (1) from the main text.

### 3.) Varshni's relation

We use Varshni's relation [2], given by $E_g(T) = E_0 - (aT^2)/(T + b)$, with $a$ and $b$ being fitting parameters, in order to extrapolate the PL's peak position at elevated temperatures from PL measurements taken below room temperature [3] (Figure S4a). The corresponding fit is shown in Figure S4b, where the dashed lines indicate error boundaries due to uncertainty in the fitting procedure. At high temperatures, the fit follows a linear redshift with -0.25 ± 0.03 meV/K.

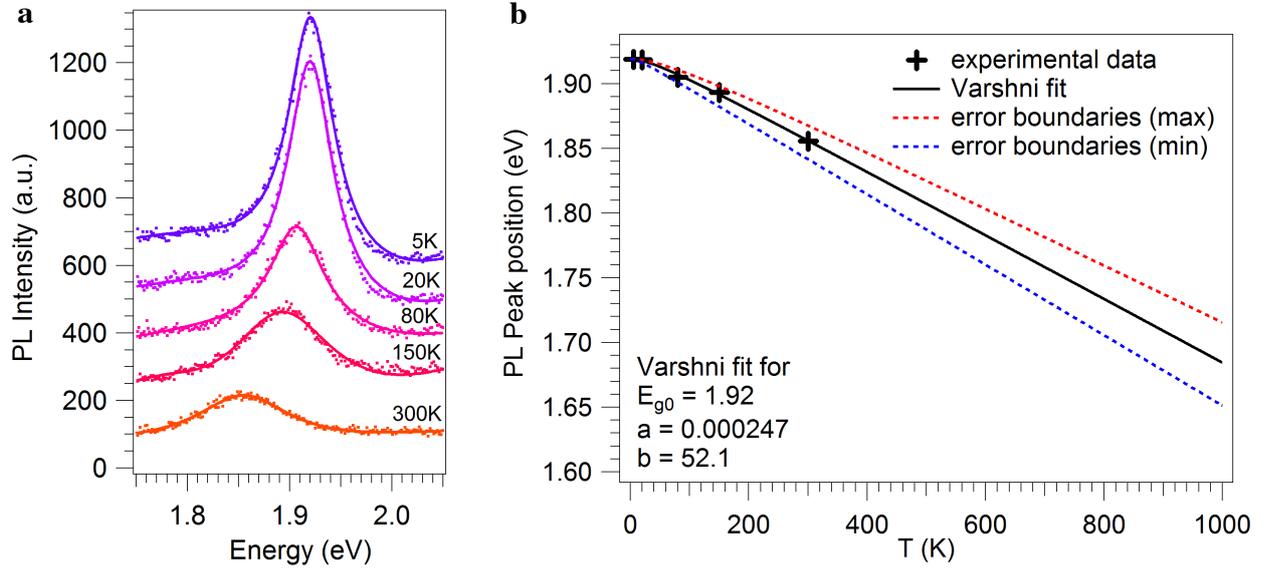

**Figure S4.** (a) Offset PL spectra at various temperatures, ranging from 5 K to 300 K. (b) Fit after Varshni's relation.

## 4.) Device modeling

IV-curves were calculated by $I_{DS} = V_{DS}/R$, with $R = 2R_C + L/(\sigma W)$, where $R_C = r_c/W$ is the contact resistance, modeled after our findings from 4-point measurements (Figure S5) at a similar device. For the simulations shown in the main article (Figures 3c & d), we slightly adapted this model (inset), while we retained the qualitative behavior (i.e. $V_{BG}$ and $V_{DS}$ dependencies) and quantitative values at high $V_{BG}$. Note, that $r_c$ rapidly decreases with increasing $V_{DS}$ and $V_{BG}$, and hence can be neglected in the region of interest, i.e. when the transistor is fully switched on and under high bias. Our model for the carrier mobility $\mu$, shown in Figure S6, considers scattering from charged impurities ($\mu_{imp} \sim T$) at low temperatures and OP phonon scattering ($\mu_{OP} \sim T^{-1.4}$) at temperatures above $\sim$200 K, as reported by Radisavljevic et al. [4], with $\mu_{80K} = 17.5$ cm$^2$V$^{-1}$s$^{-1}$ at $T = 80$ K (see Figure 1d in the main text). Further, we model the Umklapp scattering-limited thermal conductivity[4] as $\kappa = \kappa_{300}(300\ \text{K}/T)^\beta$ with $\kappa_{300} = 84\ \text{Wm}^{-1}\text{K}^{-1}$ and $\beta = 1$.

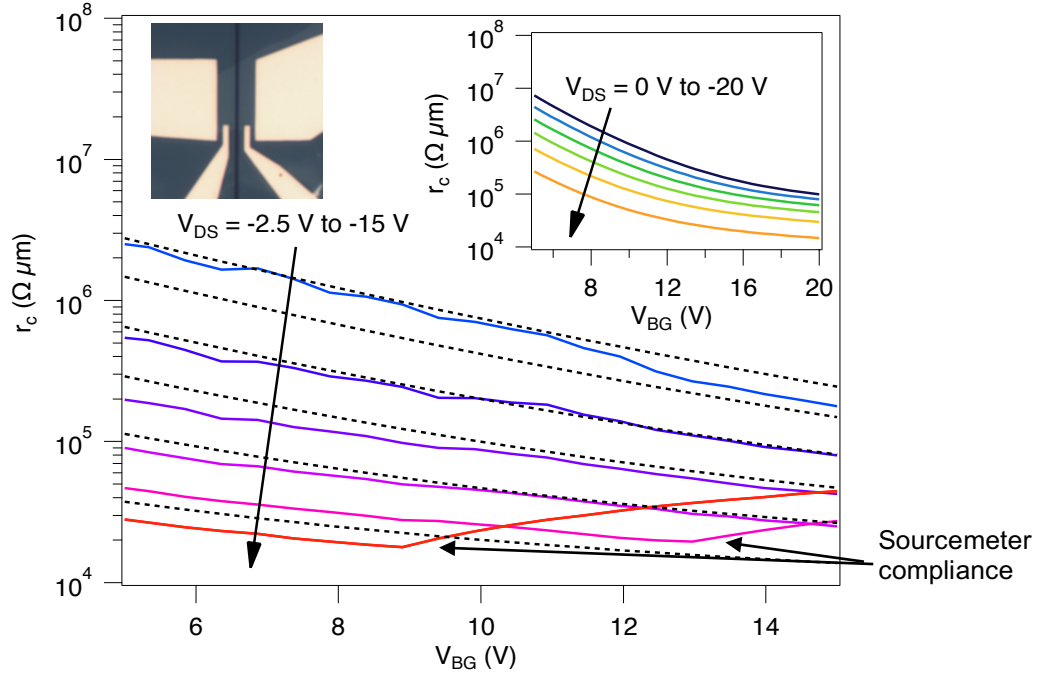

**Figure S5.** Contact resistance $r_C$ vs. $V_{BG}$ for $V_{DS}$ ranging from -2.5 V to -15 V, obtained from IV-curves were calculated by $I_{DS} = V_{DS}/R$, with $R = 2R_C + L/(\sigma W)$, where $R_C = r_c/W$ is the contact resistance, modeled after our findings from 4-point measurements (Figure S5) at a similar device (inset). 4-point measurements (device shown in the inset). The dashed lines indicate our empirical fit to the experimental data. For the device presented in the main article, we slightly adapted our model and used the data shown in the inset. Here, we model the contact resistance as $r_{C,fit} = \psi \frac{20}{1+V_{BG}}(15 \cdot 10^3 + 2.5 \cdot 10^5 \exp\left(-\frac{V_{BG}-\psi}{3}\right))$, with $\psi = 1 + \frac{20-|V_{DS}|}{4}$.

We include heat dissipation through the underlying substrate with $g_{supported} \approx 1/(L(R_{T,MoS2} + R_{T,SiN}))$, considering the thermal resistance of the $Si_3N_4$ layer $R_{T,SiN}$ and the interfacial heat resistance of the $MoS_2$ flake $R_{T,MoS2}$, while $g_{suspended} = 0$ $\text{Wm}^{-1}\text{K}^{-1}$. Here, $R_{T,SiN} = t_{SiN}/(\kappa_{SiN}A_{Surface})$ with temperature dependent $\kappa_{SiN} =$ $(\ln(T^{0.33}) - 1.176)$ Wm$^{-1}$K$^{-1}$ (fitted after Ref. 6), and $R_{T,MoS2} = 1/(hA_{Surface})$ with $h \approx 10$ MWm$^{-2}$K$^{-1}$, a typical value for van der Waals interfaces [7–9]. $A_{Surface} = LW$ depicts the interaction area. The silicon substrate is treated as heat sink at temperature $T_0$. The so-obtained thermal resistances at 80 K are $R_{T,SiN,80K} \approx 20$ kKW$^{-1}$ and $R_{T,MoS2,80K} \approx 5.5$ kKW$^{-1}$.

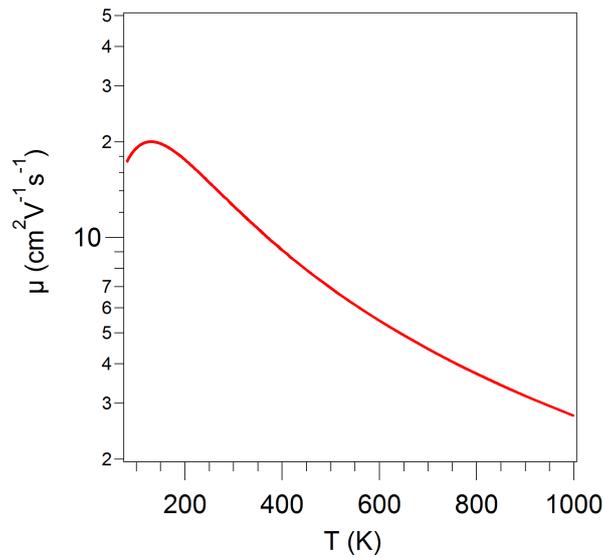

**Figure S6.** Temperature dependent mobility model, considering scattering from charged impurities at low temperatures and OP phonon scattering at temperature above ~200 K.

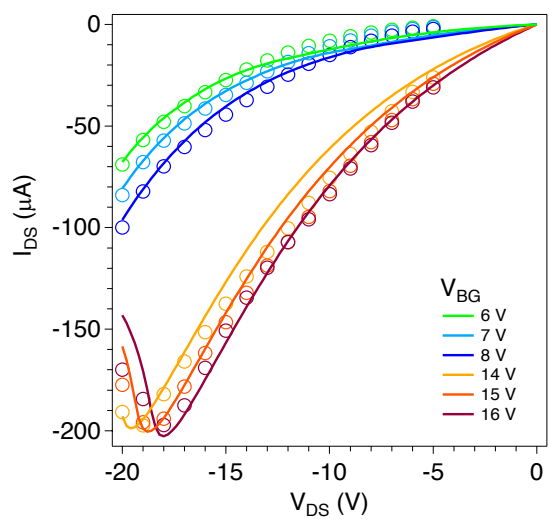

**Figure S7.** Measured (circles) and calculated (lines) IV-curves, with self-heating ($V_{BG}$ = 14 V, 15 V, 16 V) and without self-heating ($V_{BG}$ = 7 V, 8 V, 9 V)

We obtain the temperature distribution by solving the heat conduction equation (equation (2) in the main text) numerically in an iterative approach until the stepwise change in $T$ is less than 0.2 K. Figure S7 shows measured and calculated IVs for moderate biasing, where no self-heating is expected to occur, and under light emitting operation, respectively.

**5.) Influence of thermal gradient on potential**

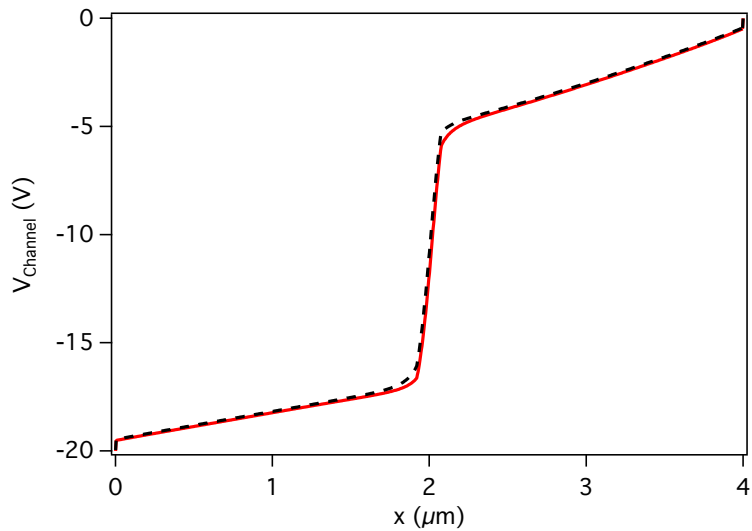

**Figure S8.** Channel potential ($V_{BG}$ = 18 V, $V_{DS}$ = -20 V) with and without thermoelectric effect.

The large thermal gradient at the trench (up to 10 K/nm) produces a thermovoltage $V_{PE} = \int S(x)\, \nabla T_e(x)\, dx$, with Seebeck coefficient $S$, that may influence the channel potential and thus the electrical transport. From our device simulations (main text) we find that the carrier concentration is high and changes only weakly from $\sim 8.4 \times 10^{12}$ cm$^{-2}$ at the drain to $\sim 4.0 \times 10^{12}$ cm$^{-2}$ at the source. $S$ is thus approximately constant along the channel, and $V_{PE} \approx S\, T_e(x)$. Further, we take the MoS$_2$ bulk value for the Seebeck coefficient ($S$ = -600 µV/K) as an upper bound, as the value for highly doped monolayers is typically lower [10, 11]. Figure S8 shows that $V_{PE}$ does not substantially alter the channel potential and can be neglected.

**6.) Upper and lower bounds for $d_{rad}$**

As demonstrated in the manuscript, the strong heat confinement over the trench ($d_{rad} \approx 50$ nm) is enabled by the low thermal conductivity of monolayer MoS$_2$ and, consequently, a short thermal transfer length $L_T$ of heat to the substrate. Here, we give upper and lower limits for $d_{rad}$.

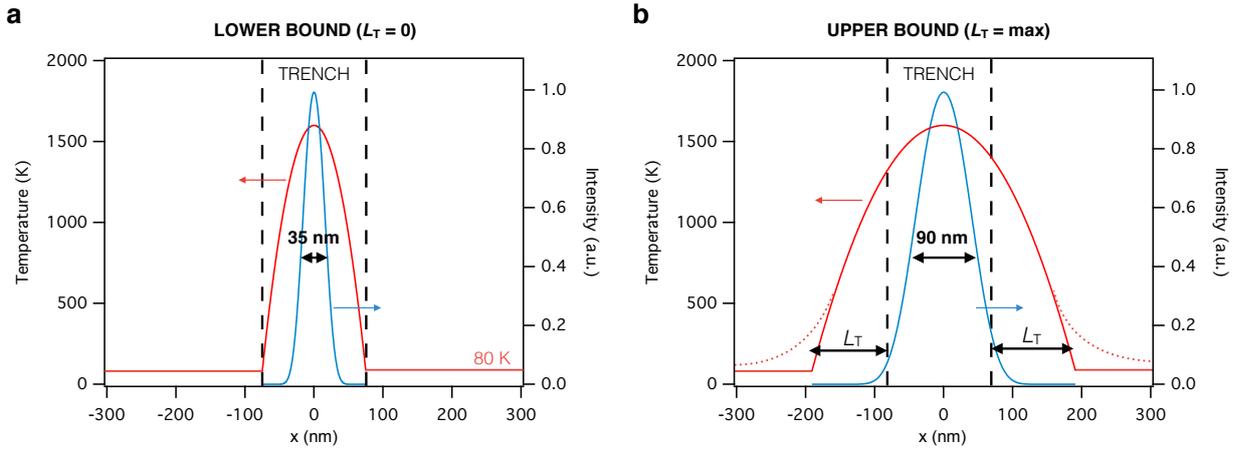

**Figure S9.** Estimation of (a) lower and (b) upper bounds for $d_{rad}$.

We start by assuming an extreme case of $L_T = 0$, i.e. the supported regions act as perfect heat sinks. In this case, the heat equation can be solved analytically (heating in the supported regions is neglected) which leads to the well-known $\propto (1 - (2x/L)^2)$ temperature distribution (*L* is the length of the trench) – see Figure S9a (red line). The resulting intensity can then be estimated from Planck's law and is shown as blue line. The so-obtained FWHM of 35 nm constitutes a lower bound for $d_{rad}$. Next, we estimate an upper bound for $L_T$ by assuming room-temperature values for all parameters: $\kappa_{MoS2} = 84$ W/Km, $\kappa_{SiN} = 0.7$ W/Km. For *h* we use 10 MW/m²K – this is a typical number for van der Waals interfaces, but chosen at the lower end of reported values (10–100 MW/m²K). Both the choice of $\kappa_{SiN}$ and that of *h* lead to thermal resistances $R_{T,MoS2}$ ($\propto 1/h$) and $R_{T,SiN}$ ($\propto 1/\kappa_{SiN}$) that overestimate the true $R_T$ values. As a result, we underestimate $g$ ($\propto 1/(R_{T,MoS2} + R_{T,SiN})$), which finally results in an upper bound for $L_T$ ($\propto \sqrt{\kappa/g}$). The so-obtained values are: $R_{T,MoS2}$ = 5500 K/W, $R_{T,SiN}$ = 8000 K/W, $g$ = 18.5 W/mK, and $L_T$ = 115 nm. Again we use the simple analytic model from above, with $L_{total} = L + 2L_T$, and plot the result as red line in Figure S9b. For the emission (blue line) we now obtain 90 nm. The true value will thus be somewhere in-between these two extreme cases, i.e. between 35 nm and 90 nm.